\documentclass[12pt,oneside]{article}
\usepackage{epsfig, cite}
\usepackage{color}
\usepackage{ifthen}
\usepackage{graphicx}

\def\p{\partial}

\def\half{{1\over 2}}
\def\({\left(}
\def\){\right)}
\def\[{\left[}
\def\]{\right]}

\def\te{\tilde{\eta}}

\def\e{\begin{equation}}
\def\q{\end{equation}}
\def\m{\begin{eqnarray}}
\def\n{\end{eqnarray}}

\begin{document}
\thispagestyle{empty} \setcounter{page}{0}
\renewcommand{\theequation}{\thesection.\arabic{equation}}

\vspace{2cm}

\begin{center}
{\Large {Running of Running of the Spectral Index \\and \\ WMAP
Three-year data}}

\vspace{1.4cm}

Qing-Guo Huang

\vspace{.2cm}

{\em School of physics, Korea Institute for
Advanced Study,} \\
{\em 207-43, Cheongryangri-Dong,
Dongdaemun-Gu, } \\
{\em Seoul 130-722, Korea}\\
\end{center}

\vspace{-.1cm}

\centerline{{\tt huangqg@kias.re.kr}} \vspace{1cm}
\centerline{ABSTRACT}
\begin{quote}
\vspace{.5cm}

Three-year data of WMAP implies not only a negative running of the
spectral index with large absolute value, but also a large positive
running of running of the spectral index with order of the magnitude
$10^{-2}$. We calculate the running of running in usual inflation
model and noncommutative inflation model. A large tensor-scalar
ratio $r\geq 1.23$ is needed in order to fit the WMAP data in the
noncommutative inflation model, which roughly saturates the
observational upper bound on it.

\end{quote}
\baselineskip18pt

\noindent

\vspace{5mm}

\newpage

\setcounter{equation}{0}
\section{Introduction}

An epoch of accelerated expansion in the early universe, inflation
\cite{infl}, dynamically resolves many cosmological puzzles in hot
big bang model, such as homogeneity, isotropy and flatness of the
universe, and generates superhorizon fluctuations without appealing
to fine-tuned initial conditions. These fluctuations become
classical after crossing out the Hubble horizon during the period of
inflation. During the deceleration phase after inflation they
re-enter the horizon, and seed the matter and the radiation
fluctuations observed in the universe. The anisotropy in CMB encodes
the information for inflation.

However, we still don't know which of the many versions of inflation
model will be picked out by the observations. An important step is
to make the observations more and more precise. Now the $\Lambda$CDM
model remains an excellent fit to the three years WMAP data and
other astronomical data \cite{wmapt}. Even though the power-law
spectral index model can fit to WMAP three-year data, the running is
slightly improved by including the small scale experiments. The
spectral index runs from $n_s>1$ at $k=0.002$Mpc$^{-1}$ to $n_s<1$
at $k=0.05$Mpc$^{-1}$.

A negative running with large absolute value would be problematic
for most inflation models, so that confirmation of this suggestive
trend is important for our understanding of early universe physics.
We can always reconstruct the potential of inflaton to get such a
running, e.g. \cite{pruns}, but the problem is how to interpret it
in a fundamental theory. On the other hand, We expect that a running
spectral index is related to the physics of the first moments of the
big bang and provides some clues into trans-Planckian physics
\cite{prunt}. In the last few years, we found that the
noncommutative effects \cite{bh,hla,hlb,hlc,pncf,ncm} always make
the power spectrum more blue and the noncommutative effects on the
small scale fluctuations can be ignored, which is nicely consistent
with observational results. There are also many explanations on the
running spectral index \cite{run,ml}.

Usually, the problem of the running spectral index is how to
construct an inflation model in a fundamental theory to provide a
negative running with large absolute value. Recently we notice that
WMAP three-year results also implies that running of the spectral
index also runs significantly from $k=0.002$Mpc$^{-1}$ to
$k=0.05$Mpc$^{-1}$, which implies a large running of running of the
spectral index with order of magnitude $10^{-2}$. This is another
problem for the running spectral index.

Our paper is organized as follows. In section 2, we give the
definition of the running of running of the spectral index and
estimate its value based on the WMAP three-year data. In section 3,
we calculate the running of running in usual slow-roll inflation
model. We investigate the modification of the running of running in
noncommutative inflation model in section 4. Section 5 contains some
concluding remarks.

\setcounter{equation}{0}
\section{Running of running of the spectral index and WMAP data}

Fit to the WMAP data is described by an eight-parameter model: four
parameters for characterizing a Friedmann-Roberton-Walker (FRW)
universe (baryonic density, matter density, Hubble constant, optical
depth), and four parameters for the primordial power spectra (the
amplitude of the power spectra, tensor-scalar ratio, the spectral
index and its running).  But here we introduce a new parameter, the
running of running, for inflation. The reason why we introduce it is
that WMAP three-year data implies a large running of running
actually.

The definition of the spectral index $n_s$ and its running
$\alpha_s$ and its running of running $\beta_s$ are respectively \m
n_s&\equiv&1+{d\ln \Delta_{\cal R}^2\over d\ln
k},\\ \alpha_s&\equiv&{dn_s\over d\ln k}, \\
\beta_s&\equiv&{d\alpha_s\over d\ln k}, \n where $\Delta_{\cal R}$
is the primordial amplitude of the power spectrum for the curvature
perturbation in comoving gauge and $k$ is the perturbation mode.

A power-law spectral index model can fit the WMAP three-year data
nicely. However the running spectral index provides a slightly
better fit to the data \cite{wmapt}. We expect that whether the
spectral index runs or not will to be distinguish in the near
future. In this paper, we focus on a running spectral index in
\cite{wmapt}. The value of the spectral index and its running and
the upper bound on the tensor-scalar ration are respectively \e
n_s=1.21^{+0.13}_{-0.16},\quad
\alpha_s=-0.102^{+0.050}_{-0.043},\quad r\leq 1.5, \q at
$k=0.002$Mpc$^{-1}$. But a red power spectrum ($n_s<1$) at
$k=0.05$Mpc$^{-1}$ is favored and the running of the spectral index
is not required at more than the $95\%$ confidence level. Thus not
only the spectral index runs, but also the running also runs. The
value of the running of running of the spectral index is estimated
at the linear approximation level as \e \label{wtrr} \beta_s\simeq
{\Delta \alpha_s\over \Delta \ln k}={0-(-0.102)\over
\ln(0.05/0.002)}\simeq 0.0318. \q A running of running of the
spectral index with the order of magnitude $10^{-2}$ is expected. To
run CAMB or CMBfast is needed if one want to get a precise result.
So the problems on the running spectral index are not only the
negative running with large absolute value, but also the large
running of running.

\setcounter{equation}{0}
\section{Running of running of the spectral index in inflation model}

In this paper, we only focus on single field inflation model. The
evolution of inflation is governed by the potential of inflaton
field. The equations of motion for an expanding universe containing
a homogeneous scalar field in a spatially flat FRW universe are \m
H^2=\({\dot a}\over a\)^2&=&{1\over 3M_p^2}\(\half {\dot \phi}^2+V(\phi) \), \label{frw} \\
\ddot \phi+3H\dot\phi+V'&=&0, \nonumber \n where $V$ is the
potential of inflaton field $\phi$ and the prime denotes the
derivative with respect to $\phi$. If ${\dot \phi}^2\ll V$ and
$|\ddot \phi|\ll 3H|\dot \phi|$, the inflaton field slowly rolls
down its potential and the equations of
motion (\ref{frw}) become \m H^2&=&{V\over 3M_p^2}, \label{sr} \\
3H\dot \phi&=&-V'. \nonumber \n To be simple, we define the
slow-roll parameters $\epsilon$ and $\eta$ as \m
\epsilon&=&{M_p^2\over 2}\({V'\over V}\)^2, \\ \eta&=&M_p^2{V''\over
V}. \n The slow-roll conditions become $\epsilon\ll 1$ and
$|\eta|\ll 1$. These slow-roll parameters also characterize the
feature of the primordial power spectrum for the perturbations.

The amplitude of the scalar power spectrum for slow-roll inflation
can be expressed as (see \cite{ll} for a review) \e \Delta_{\cal
R}^2={V/M_p^4\over 24\pi^2\epsilon}, \q and the tensor-scalar ratio
is \e r=16\epsilon. \q From slow-roll conditions, we find \e
\label{dlk} {d\over d\ln k}=-M_p^2{V'\over V}{d\over d\phi}, \q and
\m {d\epsilon\over d\ln k}&=&2\epsilon(2\epsilon-\eta),\\
{d\eta\over d\ln k}&=&-\xi+2\epsilon\eta,\\ {d\xi\over d\ln
k}&=&-\zeta-\eta\xi+4\epsilon\xi,\n where \m
\xi&\equiv&M_p^4{V'V'''\over V^2},\\
\zeta&\equiv&M_p^6{V'^2V''''\over V^3}. \n Thus the spectral index
and its running and its running of running are respectively related
to the slow-roll parameters by
\m s&=&n_s-1=-6\epsilon+2\eta,\label{cns}\\ \alpha_s&=&-24\epsilon^2+16\epsilon\eta-2\xi, \label{crun}\\
\beta_s&=&-192\epsilon^3+192\epsilon^2\eta-32\epsilon\eta^2-24\epsilon\xi+2\eta\xi+2\zeta.
\label{crrun}\n

For given spectral index $s=n_s-1$ and its running $\alpha$, solving
eq. (\ref{cns}) and (\ref{crun}), we find the running of running
becomes \e \beta_s=-\half
s\alpha_s+9\alpha_s\epsilon-4\epsilon(s^2+15s\epsilon+30\epsilon^2)+2\zeta.\label{crrunb}\q
The first term in R.H.S of (\ref{crrunb}) equals $0.01$, since
$s=0.21$ and $\alpha_s=-0.102$. The second and the third term in
R.H.S of (\ref{crrunb}) is negative, since $\epsilon>0$. In order to
get large enough running of running, $\zeta$ should not be smaller
than $0.011$. As we know, there is no such inflation model with so
large running and the running of running (for example, see
\cite{ll}).

\setcounter{equation}{0}
\section{Running of running of the spectral index in noncommutative inflation model}

To make this paper self-consistent, we briefly review the
calculation of the primordial power spectrum \cite{hlc} for
noncommutative inflation model. Then we calculate the running of
running of the spectral index and discuss fitting to WMAP three-year
results in this subsection 4.2 and 4.3.

\subsection{The primordial power spectrum and the running of running in noncommutative inflation}

Noncommutative spacetime naturally emerges in string theory
\cite{ncst}, which implies a new uncertainty relation \e
\label{urst}\Delta t_p \Delta x_p \geq l_s^2, \q where $t_p$ and
$x_p$ are the physical time and space, $l_s$ is uncertainty length
scale or string scale in string theory. Here we follow the toy model
in \cite{bh}. The spacetime noncommutative effects are encoded in a
new product among functions, namely the star product, replacing the
usual algebra product. The evolution of the background is
homogeneous and the standard cosmological equations of the inflation
does not change.

In order to make the uncertainty relationship in (\ref{urst}) more
clear in FRW background, we introduce another time coordinate $\tau$
in the noncommutative spacetime such that the metric takes the form
\e ds^2=dt^2-a^2(t)d{\vec{x}}^2=a^{-2}(\tau)d\tau^2-
a^2(\tau)d{\vec{\tau}}^2.\q Now the uncertainty relationship
(\ref{urst}) becomes \e \label{ustc}\Delta \tau \Delta x \geq l_s^2.
\q The star product can be explicitly defined as \e
\label{dstp}f(\tau,x)*g(\tau,x)=e^{-{i\over 2}l_s^2(\p_x \p_{\tau'}
- \p_{\tau} \p_{x'})}f(\tau, x)g(\tau', x')|_{\tau'=\tau, x'=x}.\q
Since the comoving curvature perturbation $\cal R$ depends on the
space and time, the equation of motion for $\cal R$ is modified by
the noncommutative effects \e \label{emsp}u_k''+\(k^2-{z_k'' \over
z_k} \)u_k=0, \q where \m \label{scaf} z_k^2(\te)&=&z^2y_k^2(\te),
\quad y_k^2=(\beta_k^+\beta_k^-)^{\half},\\ {d\te \over
d\tau}&=&\left({\beta_k^-\over \beta_k^+}\right)^\half,\quad
\beta_k^\pm =\half (a^{\pm 2}(\tau+\l_s^2k)+a^{\pm
2}(\tau-l_s^2k)),\nonumber \n here ${\cal R}_k(\te) = u_k(\te) /
z_k(\te)$ is the Fourier modes of $\cal R$ in momentum space and the
prime denotes derivative with respect to the modified conformal time
$\te$. The deviation from the commutative case encodes in
$\beta^\pm_k$ and the corrections from the noncommutative effects
can be parameterized by ${Hk\over aM_s^2}$. After a lengthy but
straightforward calculation, we get \m \label{zkt}{z_k'' \over
z_k}&=&2(aH)^2\(1+{5\over 2}\epsilon-{3\over 2}\eta-2\mu \), \\
aH&\simeq& {-1\over \te}(1+\epsilon+\mu),\nonumber \n where \e
\label{du}\mu = {H^2 k^2 / (a^2 M^4_s)}\q is the noncommutative
parameter and $M_s = l_s^{-1}$ is the noncommutative mass scale or
string mass scale. Solving eq. (\ref{emsp}) yields the amplitude of
the scalar comoving curvature fluctuations in noncommutative
spacetime \e \label{asc}\Delta_{\cal R}^2\simeq {k^3\over
2\pi^2}\left|{\cal R}_k(\te) \right|^2={V/M_p^4 \over
24\pi^2\epsilon} (1+\mu)^{-4-6\epsilon+2\eta}, \q where $H$ and $V$
take the values when the fluctuation mode $k$ crosses the Hubble
radius ($z_k''/z_k=k^2$), $k$ is the comoving Fourier mode. Using
(\ref{du}) and (\ref{dlk}), we obtain \e {d\mu \over d\ln k}\simeq
-4\epsilon\mu.\q Thus the spectral index and its running and its
running of running are respectively
\m s&=&n_s-1=-6\epsilon+2\eta+16\epsilon\mu,\label{nns}\\ \alpha_s&=&-24\epsilon^2+16\epsilon\eta-2\xi-32\epsilon\eta\mu,\label{nrun}\\
\beta_s&=&-192\epsilon^3+192\epsilon^2\eta-32\epsilon\eta^2-24\epsilon\xi+2\eta\xi+2\zeta\nonumber
\\ &+&64\epsilon\eta^2\mu-64\epsilon^2\eta\mu+32\epsilon\xi\mu. \label{nrrun}\n
The tensor-scalar ratio is still related to $\epsilon$ by
$r=16\epsilon$. Here we only sketch out the brief derivation of the
primordial power spectrum for fluctuations. When $\mu=0$, the
results in noncommutative inflation are the same as those in
commutative inflation exactly.

\subsection{Model-independent analysis}

In this subsection, we use the slow-roll parameters to make
model-independent analysis on fitting to WMAP three-year results.

First, we ignore $\xi$ and $\zeta$, there are only three parameters
$\epsilon$, $\eta$ and $\mu$. For given spectral index $n_s=1.21$,
the slow-roll parameter $\eta$ is related to $\eta$ and $\mu$ by
(\ref{nns}). Now the running and running of running are showed in
Fig. 1.
\begin{figure}[h]
\begin{center}
\leavevmode \epsfxsize=0.4\columnwidth \epsfbox{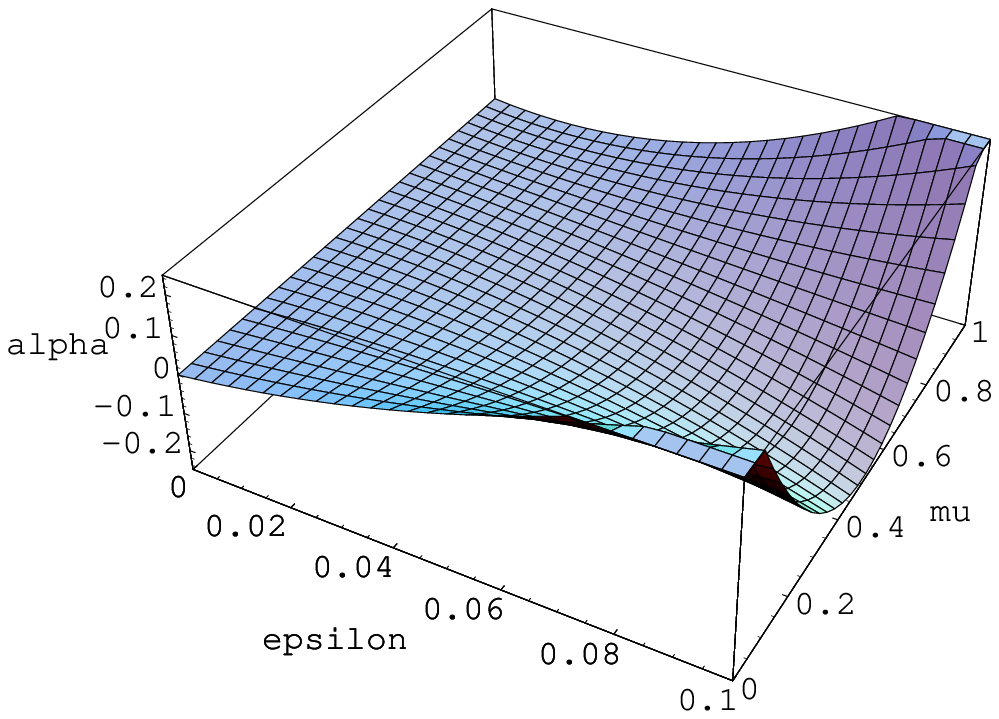}
\epsfxsize=0.4\columnwidth \epsfbox{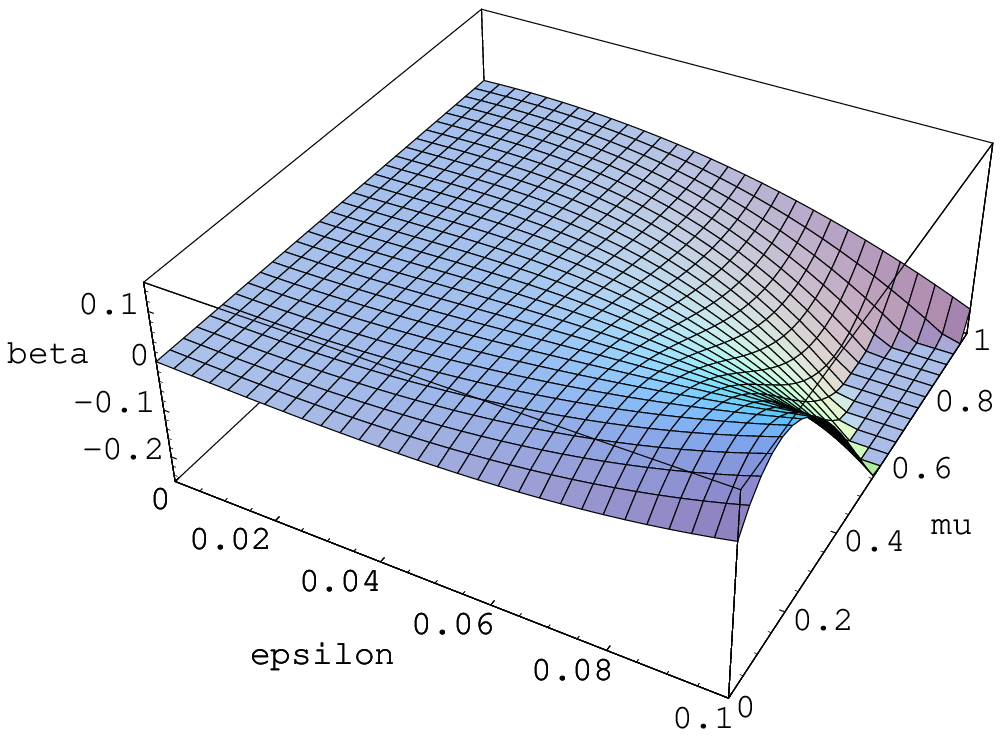}
\end{center}
\caption{Ignoring $\xi$ and $\zeta$, we shows the running and the
running of running which depend on the value of $\epsilon$ and
$\mu$, where $\eta$ is related to $\epsilon$ and $\mu$ by the
spectral index $n_s=1.21$.}
\end{figure}

If we take $n_s=1.21$, $\alpha_s=-0.102$ and $\beta_s=0.0318$ as
input, there is no free parameter. Now we find $\epsilon=0.082$,
$\eta=0.133$ and $\mu=0.33$. The tensor-scalar ratio equals
$r=1.31$. The values of parameters are reasonable and the
noncommutative inflation can fit to the data.

Next we only ignore $\zeta$. Now a new parameter $\xi$ appears and
the constraint on $\epsilon$, $\eta$ and $\mu$ is looser than
previous case. We solve $\eta$ and $\xi$ in (\ref{nns}) and
(\ref{nrun}). We find \e \xi=\half
(-\alpha_s+8s-16s\mu\epsilon+8(1-8\mu)(3-4\mu)\epsilon^2). \q Now
the running of running becomes \m \beta_s&=&-\half
s\alpha_s+9\alpha_s\epsilon-4\epsilon(s^2+15s\epsilon+30\epsilon^2)
\nonumber \\
&+&8(-\alpha_s+s^2+64\epsilon+324\epsilon^2)\epsilon\mu \nonumber\\
&-&512(s+17\epsilon)\epsilon^2\mu^2+6144\epsilon^3\mu^3.
\label{nrrunb}\n For given spectral index $n_s=1.21$ and its running
$\alpha_s=-0.102$, the running of running is showed in Fig. 2.
\begin{figure}[h]
\begin{center}
\leavevmode \epsfxsize=0.4\columnwidth \epsfbox{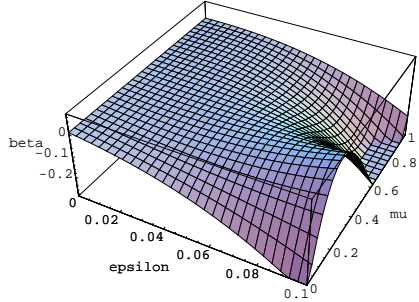}
\end{center}
\caption{Ignoring $\zeta$, we show the running of running which
depends on the value of $\epsilon$ and $\mu$, where $\eta$ and $\xi$
are related to $\epsilon$ and $\mu$ by the spectral index $n_s=1.21$
and the running of the spectral index $\alpha_s=-0.102$.}
\end{figure}

For ignoring $\zeta$, there are four free parameters, $\epsilon$,
$\eta$, $\xi$ and $\mu$. However there are only three constraints. A
degree of freedom is left. We show the constraint on the $\epsilon$,
$\eta$ and $\mu$ in Fig. 3.
\begin{figure}[h]
\begin{center}
\leavevmode \epsfxsize=0.4\columnwidth \epsfbox{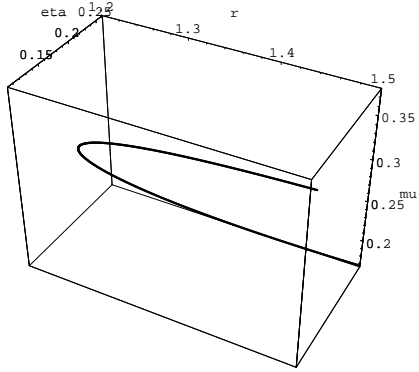}
\end{center}
\caption{Ignoring $\zeta$, we show the constraint on $r=16\epsilon$,
$\eta$ and $\mu$ for $n_s=1.21$, $\alpha_s=-0.102$ and
$\beta_s=0.0318$.}
\end{figure}
According to this figure, we find a large tensor-scalar ratio $r\geq
1.23$ is needed. On the other hand, the amplitude of the power
spectrum is $\Delta_{\cal R}=2.09\times 10^{-9}$ \cite{wmapt}. Using
(\ref{asc}), we find the Hubble parameter during the period of
inflation is roughly $\sqrt{\pi^2\Delta_{\cal R}^2r\over 2}M_p\simeq
10^{-4}M_p\simeq 2.4\times 10^{14}$Gev. Since $\mu\sim
H^4/M_s^4\simeq 0.3$, the noncommutative scale or string scale is
roughly $3.2\times 10^{14}$Gev, which is lower than GUT scale.

The constraints for WMAP+SDSS is similar to WMAP+2dFGRS. For the
case with tensor perturbations, WMAP+SDSS gives a more stringent
constraint on the tensor-scalar ratio as \e r\leq 0.67\quad
\hbox{WMAP+SDSS}, \quad r\leq 1.0\quad \hbox{WMAP+2df}, \q at $95\%$
CL. If so, the noncommutative inflation cannot provide a running
with large enough absolute value within reasonable parameters space.
On the other hand, if we ignore the tensor perturbations, the
spectral index and its running are respectively \cite{sdss} \e
n_s=0.895^{+0.041}_{-0.042}, \quad \alpha_s=-0.040^{+0.027}_{-0.027}
\q for WMAP+SDSS. Now the value of running of running is roughly
$\beta_s\simeq 0.0124$. Setting $\xi=\zeta=0$, we find
$\epsilon=0.081$, $\eta=0.117$, $\mu=0.113$ and the tensor-scalar
ratio is $r=1.3$. Taking $\xi$ into account, we show the fit to the
running spectral index in fig. 4 and we find $r\geq 1.25$ is needed.
\begin{figure}[h]
\begin{center}
\leavevmode \epsfxsize=0.4\columnwidth \epsfbox{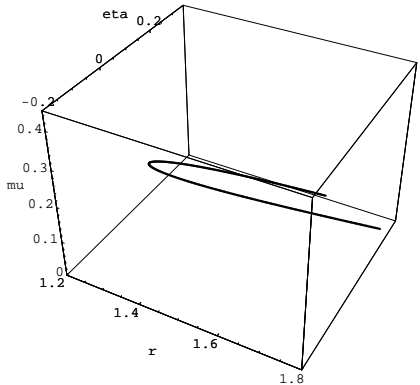}
\end{center}
\caption{Ignoring $\zeta$, we show the constraint on $r=16\epsilon$,
$\eta$ and $\mu$ for $n_s=0.895$, $\alpha_s=-0.04$ and
$\beta_s=0.0124$.}
\end{figure}
Considering the upper bound on the tensor-scalar ration given by
WMAP group, we conclude that the noncommutative cannot provide a
good fit to WMAP+SDSS data.

\subsection{Check some typical inflation models}

In this subsection we check two inflation models: power-law
inflation model and chaotic inflation model. The small field
inflation models predict a small amplitude of tensor perturbations
and the corrections from the spacetime noncommutative effects can be
neglected.

Power-law inflation is governed by the potential of inflaton with \e
V=V_0\exp\(-\sqrt{2\over p}{\phi\over M_p}\). \q The slow-roll
parameters are \e \epsilon=1/p, \quad \eta=2/p, \quad
\xi=4/p^2,\quad \zeta=8/p^3. \q The spectral index and its running
and its running of running are respectively \m s&=&n_s-1=-{2\over
p}+{16\over p}\mu, \\ \alpha_s&=&-{64\over p^2}\mu, \\
\beta_s&=&{256\over p^3}\mu. \n In commutative spacetime, $\mu=0$,
the spectral index does not run at all. In noncommutative case,
there are only two parameters, $p$ and $\mu$. Requiring $n_s=1.21$
and $\alpha_s=-0.102$ yields $p=13.9$ and $\mu=0.307$ and thus
$\beta_s=0.0293$, which is slightly smaller than that from data.

Chaotic inflation is dominated by the inflaton potential with \e
V={\lambda\over p}\phi^p. \q The value of inflaton at the time
corresponding to number of e-folds $N$ before the end of inflation
is given by \e \phi_N=\sqrt{2pN}M_p. \q The slow-roll parameters are
expressed as \m \epsilon={p\over 4N},\quad \eta&=&{p-1\over
2N},\quad \xi={(p-1)(p-2)\over 4N^2},\\
\zeta&=&{(p-1)(p-2)(p-3)\over 8N^3}. \n The spectral index and its
running and its running of running are
\m s&=&-{p+2\over 2N}+{4p\over N}\mu, \\
\alpha_s&=&-{p+2\over 2N^2}-{4p(p-1)\over N^2}\mu,\\
\beta_s&=&-{p+2\over N^3}+{4p(p-1)(p-2)\over N^3}\mu. \n For N=50,
requiring $n_s=1.21$ and $\alpha_s=-0.102$ yields $p=13.9$ and
$\mu=0.233$ and thus $\beta_s=0.0158$. For $n_s=1.21$ and
$\alpha_s=-0.102$, we scan the range with $N\in [47,61]$ and show
the running of running in Fig. 4.

\begin{figure}[h]
\begin{center}
\leavevmode \epsfxsize=0.4\columnwidth \epsfbox{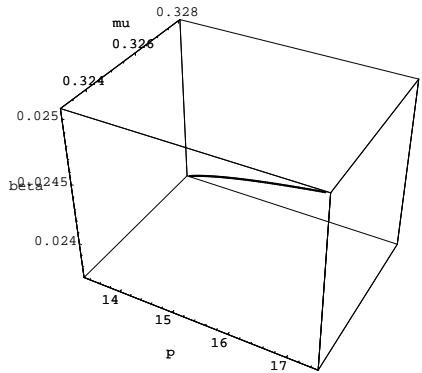}
\end{center}
\caption{The line corresponds to $n_s=1.21$, $\alpha_s=-0.102$ for
$N\in [47,61]$.}
\end{figure}

However, we can not find a reasonable solution for $n_s=1.21$,
$\alpha_s=-0.102$ and $\beta_s=0.0318$. The value of running of
running is roughly $0.0245$ which is not large enough to fit the
data.

\setcounter{equation}{0}
\section{Conclusion}

In this paper we point out that there is another problem for the
running spectral index, i.e. the running of the spectral index also
runs. It is a new challenge for  inflation model building. However
noncommutative inflation model still provide a nice explanation on
the running of running. But a large amplitude of the tensor
perturbations is needed. Noncommutative power-law inflation is still
nice, but the noncommutative chaotic inflation can not provide a
large enough running of running.

If we take the running of running into account, the tensor-scalar
ratio roughly saturates the upper bound on it in WMAP three-year
data and the noncommutative scale or string scale is roughly
$3.2\times 10^{14}$Gev. However, noncommutative inflation model
cannot provide a nice fit to WMAP+SDSS data. We expect that the
observations, e.g. more years data of WMAP and Planck, can confirm
or rule it out in near future at high level of confidence.

\vspace{.5cm}

\noindent {\bf Acknowledgments}

We would like to thank P. Chingangbam for useful discussions.

\end{document}